\documentclass{emulateapj}
\usepackage{apjfonts}
\usepackage{amsmath}
\usepackage{natbib}          
\shorttitle{X-RAY FLARES FROM LMXBS IN NGC 4697}
\shortauthors{SIVAKOFF, SARAZIN, \& JORD\'{A}N}
\slugcomment{Astrophysical Journal Letters, submitted}

\begin{document}

\title{Luminous X-ray Flares from Low Mass X-ray Binary Candidates
in the Early-Type Galaxy NGC 4697}

\author{
Gregory R. Sivakoff\altaffilmark{1},
Craig L. Sarazin\altaffilmark{1}, and
Andr\'{e}s Jord\'{a}n\altaffilmark{2,3}
}

\altaffiltext{1}{Department of Astronomy, University of Virginia,
P. O. Box 3818, Charlottesville, VA 22903-0818;
grs8g@virginia.edu, sarazin@virginia.edu}
\altaffiltext{2}{European Southern Observatory,
Karl-Schwarzschild-Str.\ 2
85748 Garching bei M\"{u}nchen, Germany; ajordan@eso.org}
\altaffiltext{3}{Astrophysics, Denys Wilkinson Building, University of Oxford,
1 Keble Road, Oxford, OX1 3RH, UK}

\begin{abstract}
We report results of the first search specifically
targeting short-timescale X-ray flares from low-mass X-ray binaries
in an
early-type galaxy.
A new method for flare detection is presented.
In NGC 4697, the nearest, optically luminous, X-ray faint elliptical
galaxy, 3 out of 157 sources are found to display flares at $>99.95 \%$
probability, and all show more than one flare.
Two sources are coincident with globular clusters
and show flare durations and luminosities
similar to (but larger than) Type-I X-ray superbursts found in
Galactic neutron star (NS) X-ray binaries (XRBs).
The third source shows more extreme flares.
Its flare luminosity ($\sim 6 \times 10^{39}$ erg s$^{-1}$)
is very super-Eddington for an NS and 
is similar to the
peak luminosities of the brightest Galactic black hole (BH) XRBs.
However, the flare duration ($\sim 70$ s) is much
shorter than are typically seen for outbursts reaching those luminosities
in Galactic BH sources.
Alternative models for the flares are considered.
\end{abstract}


\section{Introduction} \label{sec:flare_intro}

Galactic X-ray binaries exhibit a wide-range of short-timescale
variability.
X-ray bursts from such binaries have been studied since the
mid-1970's \citep[e.g.,][]{GGS+1976,BCE+1976}.
Type-I bursts are found in low-mass
X-ray binaries (LMXBs), and are thought to be due to thermonuclear
flashes on the neutron-star (NS) surface with maximum burst
luminosities at the Eddington limit.
Eight bursts with hours long durations,
but otherwise similar to Type-I bursts have been observed
\citep{K2004}. These superbursts are relatively rare and are
thought to have recurrence timescales on the order of a year.

Short-timescale variability has also been seen in high-mass X-ray
binaries (HMXBs).
LMC X$-$4, an HMXB-pulsar,
has a persistent luminosity of
$\sim 2 \times 10 ^{38} \, {\rm erg\,s}^{-1}$ and super-Eddington
flares (up to $\sim 2 \times 10 ^{39} \, {\rm erg\,s}^{-1}$)
with durations of $\sim70$ s \citep{MEW2003}.

Since black holes (BHs) lack a surface, Type-I bursts do not occur.
Typical short flares observed in BH-XRBs tend to last on the order of
days \citep{MR2005} as opposed to minutes as in NS-XRBs; however,
rapid flares have been seen
\citep[e.g.,][]{GMR1996,WK2000}.
In GRS1915+105, which has been
relatively luminous ($\sim 3 \times 10 ^{38} \, {\rm erg\,s}^{-1}$)
since its discovery, these rapid flares typically
last for minutes or less, have peak-to-persistent flux ratios
$\lesssim 10$,
and often recur on both regular and irregular intervals
of minutes or less
\citep{YRA+1999}.
V4641 Sgr, an HMXB, showed variations
of a factor up to 500 on the timescale of minutes
\citep{WK2000}.

In this Paper, we perform the first search specifically targeting
short-timescale flares from LMXBs in an early-type galaxy.
Since nearby early-type galaxies have many bright LMXBs
\citep[$> 10 ^{37}$ erg s$^{-1}$; e.g.,][]{SIB2000},
they have a high
potential for having sources that exhibit interesting flare behaviors.
We targeted NGC 4697, the nearest
\citep[$11.7 \, {\rm Mpc}$,][]{TDB+2001}
optically luminous ($M_B < -20$) elliptical galaxy.
As an X-ray faint galaxy (low $L_{\rm X}/L_{\rm B}$),
most of its X-ray emission is resolved into 158 point sources
(Sivakoff et al.\ 2005, in preparation, hereafter Paper IV).
Five {\it Chandra} observations of NGC 4697 give extended time
coverage for searching for variability.
We discuss a new method for detecting flares in the low count
regime, and characterize short timescale flares in three LMXBs.
All errors listed are for 90\% confidence intervals, and all
count-rates and fluxes refer to the
0.3--10 keV band.

\section{Observations and Data Reduction} \label{sec:flare_n4697obs}

{\it Chandra} has observed NGC 4697 five times (observations 0784,
4727, 4728, 4729, and 4730), 2000 January 15, 2003 December 26,
2004 January 06, February 02, and August 18 using the Advanced
CCD Imaging Spectrometer (ACIS) S3 detector.
After removal of background flares, the ontimes were
37651, 40447, 36072, 32462, and $40574 \, {\rm s}$.
All observations were analyzed under {\sc ciao 3.1}
with {\sc caldb 2.28}.
We detected 158 sources
using the wavelet detection algorithm ({\sc ciao wavdetect} program).
Optical identifications were used to refine the coordinates to
$0\farcs4$ accuracy, with larger errors for
fainter sources away from the chip center.
More details concerning the X-ray
observations, data analysis, and source properties of the whole sample
are given in
Paper IV.

We observed the center of NGC 4697 with {\it Hubble Space Telescope
Advance Camera for Surveys} (HST-ACS),
acquiring
two $375 \, {\rm s}$ exposures in the F475W (g475) band, two
$560 \, {\rm s}$ exposures in the F850LP (z850) band, and one
$90 \, {\rm s}$ F850LP exposure. Source detection and characterization
were performed similarly to \citet{JBP+2004}, leading to a list of 
globular clusters (GCs)
and other optical sources.
Details concerning the HST-ACS
observation, data analysis, and optical source properties
are given in Jord\'{a}n et al.\ (2005, in preparation).

\section{Detection of Flaring Sources} \label{sec:flare_detection}

\tabletypesize{\scriptsize}
\begin{deluxetable*}{lcccccccccc}
\tablecaption{Observed Properties of Short Timescale Flares in NGC 4697 \label{tab:n4697_flare_obs}}
\tablehead{
Source Name&
R.A.&
Dec.&
Obs.&
$N$ &
$n$ &
$\Delta t$ &
$t_0$ &
$P_{\rm constant} $&
$P_{\rm constant}^\prime $&
$P_{\rm constant, joint}$\\
&
(h m s)&
($\arcdeg$ $\arcmin$ $\arcsec$)&
&
&
&
(s)&
(MJD)&
(\%)&
(\%)&
(\%)\\
(1)&
(2)&
(3)&
(4)&
(5)&
(6)&
(7)&
(8)&
(9)&
(10)&
(11)
}
\startdata
CXOU J124837.8$-$054652 (Src.\ A)&12 48 37.86&$-$05 46 52.9&0784        &14&5&      1047&51558.79424&0.700&1.787&\\
                       &           &             &4727        &16&4&   \phn628&52999.75883&2.729&7.006&\\
                       &           &             &4728        &14&4&   \phn509&53010.70296&1.212&3.096&\\
                       &           &             &0784,4727--8&  & &          &           &     &     &0.039\\
CXOU J124831.0$-$054828 (Src.\ B)&12 48 31.04&$-$05 48 28.8&4727        & 9&5&      1329&53000.04327&0.224&0.552&\\
                       &           &             &4728        & 9&4&      1420&53010.72445&3.368&7.982&\\
                       &           &             &4729        & 6&5&      5654&53047.59917&4.327&9.007&\\
                       &           &             &4727--8     &  & &          &           &     &     &0.441\\ 
                       &           &             &4727--9     &  & &          &           &     &     &0.040\\
CXOU J124839.0$-$054750 (Src.\ C)&12 48 39.03&$-$05 47 50.2&0784        &20&4&\phn\phn68&51558.95771&0.013&0.034&\\
                       &           &             &4727        &20&3&\phn\phn50&53000.05857&0.547&1.420&\\
                       &           &             &0784,4727   &  & &          &           &     &     &0.005
\enddata
\end{deluxetable*}

A bright LMXB ($10 ^{38} \, {\rm erg\,s}^{-1}$) will have
only $\sim 25$ counts in a $\sim 40 \, {\rm ks}$ {\it Chandra} observation of
NGC 4697, making it difficult to search for intra-observation variability.
Tests which require binning the events and the Kolmogoroff-Smirnov test
are not useful for detecting flares with small numbers of events.
Instead, we consider arrival times of individual events and
compare them to a constant rate Poisson distribution.
We searched for
the shortest time $t_{n}$ over which $n$ consecutive photons
(hereafter an $n$-tuple) arrived.
Assuming there are $N$ photons in the entire ontime
($t_{\rm exp}$), the probability that at least $n$ photons are
seen in $t_{n}$ due to Poisson fluctuations is
given by the incomplete gamma function,
\begin{equation}
\label{eq:prob_single}
P(n \ {\rm in} \ t_{n} \ {\rm given}\  N \ {\rm in} \ t_{\rm exp}) =
\frac{1}{\Gamma(n-1)} \int^{N t_{n} / t_{\rm exp}}_{0}
e^{-a} a^{n-2} da \, .
\end{equation}
Since we search $N - n + 1$ $n$-tuples, the probability that the $n$ observed
photons that are seen in $t_{n}$ are not due to random fluctuations
is given by 
\begin{equation}
\label{eq:prob_ntuple}
P_{\rm flare} (n) =
[1-P(n \ {\rm in} \ t_{n} \ {\rm given}\  N \ {\rm in} \ t_{N})] ^{N - n + 1}
\, ,
\end{equation}
assuming the $n$-tuples are independent.
The probability that the flare resulted from a statistical fluctuation is
$ P_{\rm constant} (n) = 1 - P_{\rm flare} (n)$.

With each observation, we search through $N-1$ different sets of $n$-tuples,
i.e., $n = 2 \ldots N$.
Thus, eq.~\ref{eq:prob_ntuple} overestimates the probability a detected
flare of $n$ photons is real.
However, the $N-1$ sets of $n$-tuples are not independent.
Adding probabilities of producing a flare from a random
fluctuation would underestimate the probability that the flare was real.
To determine the probabilities more accurately, we ran sets of 
$>200,000$ Monte Carlo simulations assuming the source emitted at
a constant rate, and we determined the number of false flares
detected by our algorithm due to statistical fluctuations.
The simulations, which included the effects of finite frame times, read-out
time, and background, were analyzed identically to the actual observations.
From simulations, we derived a correction factor to
eq.~(\ref{eq:prob_ntuple}) to give the correct probability that
a flare was real, $P_{\rm flare}^\prime$.
We determined the ratio between $P_{\rm constant}^\prime$ and
$P_{\rm constant}$ when $P_{\rm constant} < 5\%$;
this ratio
$ A(N) \equiv P_{\rm constant}^\prime / P_{\rm constant}$
depended only on the mean number of counts $N$ for a set of simulations
and was typically 2--3.
We linearly
interpolated between simulation sets with different N to find
$P_{\rm constant}^\prime$ for each flare.
After finding a flaring source, we checked all $n-$tuples to see if
there were other less probable but still significant flares in the
same observation.
We did not find multiple flares in the same observation.
As a caveat, note that the use of the total
observed counts in eq.~(\ref{eq:prob_single}) would create a bias
against detecting multiple intra-observation flares.

We found that several of the sources showed flares in different
observations with similar properties.
Detecting multiple flares from a source makes it even less
likely that the flares are due to statistical fluctuations.
Let $P_{\rm constant, i}^\prime $ be the probability that the most significant
flare within a given observation $i$ is due to a statistical fluctuation.
When $P_{\rm constant, i} > 5\%$, we
conservatively set $P_{\rm constant, i}^\prime=1$.
We only consider sequences of flares with properties that are all identical
within the errors.
With five observations, the joint probability that the flare sequence
is a statistical fluctuation is,
\begin{equation}
\label{eq:prob_flare}
P_{\rm constant, joint} = \binom{5}{k} \prod_{i = 1}^{5}
\, P^\prime_{{\rm constant,} i}
\, ,
\end{equation}
where $k$ is the number of observations with $P_{\rm constant, i} < 5\%$.

Whenever $P_{\rm constant, joint} < 0.32\%$, we considered a source
to be a flaring source.
At this level, we expect $< 0.5$
sources with apparent flares due to random fluctuations;
in actuality,
we find four sources with significant flares, all of
which have joint probabilities more consistent with $<0.1$ sources being
due to random fluctuations.

\tabletypesize{\scriptsize}
\begin{deluxetable*}{lccccccc}
\tablecaption{Inferred Properties of Short Timescale Flares in NGC 4697 \label{tab:n4697_flare_inf}}
\tablehead{
Source Name&
Obs.&
$n_{\rm flat}$&
$r_{\rm flat}$&
$\Delta t_{\rm flat}$&
$r_{\rm flat}$/$r_{\rm persistent}$&
$L_{\rm flare}$&
$E_{\rm flare}$\\
&
&
&
($10^{-3}$ s$^{-1}$)&
(s)&
&
($10^{38} \,{\rm erg\,s}^{-1}$) &
($10^{41} \, {\rm erg}$) \\
(1)&
(2)&
(3)&
(4)&
(5)&
(6)&
(7)&
(8)
}
\startdata
CXOU J124837.8$-$054652 (Src.\ A)&0784        &$3.7^{+4.1}_{-2.2}$&$\phn3.2^{+\phn17.4}_{-\phn1.2}$&$      1168^{+      1418}_{-      1038}$&$14^{+    \phn\phn99}_{-\phn\phn\phn2}$&$\phn4.2^{+\phn23.2}_{-\phn\phn1.5}$&$4.9^{+5.4}_{-3.0}$\\
                       &4727        &$2.6^{+3.4}_{-1.7}$&$\phn3.5^{+\phn36.5}_{-\phn2.3}$&$   \phn754^{+      3573}_{-   \phn701}$&$12^{+       \phn161}_{-\phn\phn\phn6}$&$\phn5.8^{+\phn60.7}_{-\phn\phn3.9}$&$4.4^{+5.6}_{-2.9}$\\
                       &4728        &$2.3^{+3.0}_{-1.5}$&$\phn4.5^{+\phn24.8}_{-\phn2.6}$&$   \phn609^{+      1681}_{-   \phn534}$&$17^{+       \phn117}_{-\phn\phn\phn6}$&$\phn7.3^{+\phn39.8}_{-\phn\phn4.2}$&$3.7^{+4.8}_{-2.4}$\\
                       &0784,4727--8&                   &                                &$   \phn844^{+      1399}_{-   \phn454}$&$14^{+    \phn\phn74}_{-\phn\phn\phn3}$&$\phn5.8^{+\phn25.4}_{-\phn\phn2.0}$&$4.3^{+3.0}_{-1.6}$\\
CXOU J124831.0$-$054828 (Src.\ B)&4728        &$4.2^{+4.6}_{-2.6}$&$\phn2.8^{+\phn11.7}_{-\phn0.5}$&$      1483^{+   \phn789}_{-      1244}$&$26^{+       \phn234}_{-\phn\phn\phn2}$&$\phn4.7^{+\phn19.5}_{-\phn\phn0.9}$&$7.0^{+7.7}_{-4.2}$\\
                       &4729        &$3.0^{+3.9}_{-2.2}$&$\phn1.8^{+\phn43.5}_{-\phn0.8}$&$      1662^{+      2739}_{-      1621}$&$13^{+          1274}_{-\phn\phn\phn2}$&$\phn2.9^{+\phn69.7}_{-\phn\phn0.8}$&$4.8^{+6.2}_{-3.5}$\\
                       &4730        &$3.8^{+4.2}_{-2.3}$&$\phn0.6^{+\phn19.0}_{-\phn0.1}$&$      6320^{+      1067}_{-      6235}$&$22^{+\phn\phn\infty}_{-\phn\phn\phn1}$&$\phn1.0^{+\phn31.1}_{-\phn\phn0.1}$&$6.2^{+6.8}_{-3.8}$\\
                       &4727--8     &                   &                                &$      1573^{+      1425}_{-      1022}$&$20^{+       \phn649}_{-\phn\phn\phn1}$&$\phn3.8^{+\phn18.4}_{-\phn\phn0.9}$&$5.9^{+4.9}_{-2.7}$\\
                       &4727--9     &                   &                                &$      3155^{+      1015}_{-      2187}$&$20^{+\phn\phn\infty}_{-\phn\phn\phn1}$&$\phn2.9^{+\phn26.3}_{-\phn\phn0.5}$&$6.0^{+4.0}_{-2.2}$\\
CXOU J124839.0$-$054750 (Src.\ C)&0784        &$2.8^{+3.7}_{-1.9}$&$   36.5^{+\phn85.4}_{-   10.9}$&$\phn\phn78^{+\phn\phn74}_{-\phn\phn52}$&$76^{+       \phn220}_{-   \phn\phn20}$&$   48.5^{+   113.5}_{-   \phn14.5}$&$3.8^{+4.9}_{-2.5}$\\
                       &4727        &$2.1^{+3.3}_{-1.5}$&$   37.1^{+   116.3}_{-   26.0}$&$\phn\phn57^{+   \phn226}_{-\phn\phn42}$&$96^{+       \phn382}_{-   \phn\phn63}$&$   61.8^{+   193.5}_{-   \phn43.3}$&$3.5^{+5.5}_{-2.5}$\\
                       &0784,4727   &                   &                                &$\phn\phn68^{+   \phn119}_{-\phn\phn33}$&$86^{+       \phn220}_{-   \phn\phn33}$&$   55.2^{+   112.2}_{-   \phn22.8}$&$3.7^{+3.7}_{-1.8}$
\enddata
\end{deluxetable*}

One of the significant flaring sources is
CXOU J124836.3$-$055333, one of the brightest objects in our
sample
($\sim 5$--$9 \times 10^{-14} \, \rm{erg\,s}^{-1}\,{\rm cm}^{-2}$).
This source is within $0\farcs5$ of a Galactic star, Tycho-2 4955-1175-1,
and is very likely to be associated with this star;
we exclude it from further discussion
since it is unlikely to be an LMXB in NGC 4697.

In Table~\ref{tab:n4697_flare_obs}, we list observed properties of
the significant flaring
sources associated with NGC 4697.
Columns 1--4 list
the source name,
position,
and
observation number in which the flare occurred.
Columns 5--8 list
the total number of events $N$ from the source during this observation, 
the number of events in the flare $n$,
the flare duration $\Delta t$,
and the Modified Julian Day (MJD) of the first event, $t_0$.
Column 9 lists the probability that a given $n$-tuple is a statistical
fluctuation, while
column 10 gives the probability including all of the $n$-tuples which were
searched.
Column 11 gives the joint probability that a sequence of similar flares
during several observations are all just statistical fluctuations.

For each event in a flaring source, we examined the chip coordinates
and {\it ASCA} grades to eliminate previously unknown flickering pixels
as possible false positive flares.
Although the four flare events in Obs.\ 0784 of Src.\ C occur at only two chip positions, their ASCA grades
(2 \& 0, 6 \& 2) indicate they are unlikely to be due to flickering
pixels.
We note that the time intervals between events at the same chip positions
in Obs.\ 0784 of Src.\ C are $6.4 \, {\rm s}$ and
$12.8 \, {\rm s}$, much less than the aspect motion
periods.
Thus, it is likely that subsequent events from the same source
can occur at the same detector position.
The other sources, which all
had longer flare durations, did not have duplicate chip
positions in their flares.

It is likely that the observed flare duration underestimates the
actual duration due to the small number of photons detected, and
that other properties may be biased by the flare detection algorithm.
On the other hand, if a photon due to emission outside of the flare duration
arrives just before or after a flare, this could lead to an
overestimate of the flare properties.
To assess these effects, we adopted a simple model for the temporal
development of the flare, and used Monte Carlo simulations to derive the
best-fit model flare parameters and their uncertainties.
For the flare, we adopted a top-hat model, characterized by a constant
flare rate, $r_{\rm flat}$, over a burst duration,
$\Delta t_{\rm flat}$, beginning at a given time, $t_{\rm
0,flat}$.
Outside of the flare duration, we assumed a constant persistent source rate
$r_{\rm persistent}$.
For a given model flare rate $r_{\rm flat}$,
duration $\Delta t_{\rm flat}$,
and
beginning time $t_{ \rm 0,flat}$,
we performed 200,000 Monte Carlo simulations.
We varied the model parameters until we found the model which was
most likely to have reproduced the observed flare properties.
For this best-fit model, the symmetric 90\% confidence region for each
model parameter was determined from the simulations.
The rates and durations were used to calculate X-ray luminosities
and fluences for each observation using conversion factors
based on the best-fit model for the cumulative X-ray spectrum of
all the sources, a power-law with index of 1.47
(Paper IV).
The spectral model was corrected for Galactic absorption and 
the QE degradation for each observation.

In Table~\ref{tab:n4697_flare_inf}, we list the best-fit model properties
of the flaring sources.
Columns 1--2 lists the source name and observation number.
Columns 3--8 list the best-fit flare properties:
number of flare photons ($n_{\rm flat}$),
flare rate ($r_{\rm flat}$),
flare duration ($\Delta t_{\rm flat}$),
ratio of flare rate to persistent rate,
flare luminosity ($L_{\rm flat}$),
and flare fluence ($E_{\rm flat}$).
Rows with multiple observations listed display the averages of
the flare duration, ratio of flare rate to persistent rate, flare
luminosity, and flare fluence of the matching flares.

Given our limited temporal coverage, it is difficult to place strong limits
on the recurrence timescale ($\Delta T$) of the flares.
Our simulations show that the probability that our algorithm will detect
a flare is only $\sim 50\%$.
However, the fact that two flares were not seen in any single observation
and that there are observations without observed flares for each of the
sources implies that $ \Delta T \gtrsim 5 \, {\rm hr}$.
One upper limit on $\Delta T$ is given by the shortest observed
time between observations with flares;
this is 11 days for 
Src.\ A and B, and
1441 days for Src.\ C.
However, detections of flares in two or three out of five
observations lasting $\sim 10 \, {\rm hr}$ suggest that the recurrence time
is much shorter than this, $\Delta T \approx 10 \, {\rm hr}$.
One caveat is that it is possible that most of the LMXBs in NGC 4697
are undergoing flares with a longer recurrence timescale, and we have
just selected the three sources where flares occurred within several of
our observing windows.

\section{Discussion} \label{sec:flare_discussion}

\subsection{CXOU J124837.8$-$054652 (Src.\ A) \& CXOU J124831.0$-$054828
(Src.\ B):
Type-I X-ray Superbursts?} 

Sources A and B
both show similar flares,
with average durations of $844^{+1399}_{-\phn454} \, {\rm s}$ and
$3155^{+1015}_{-2187} \, {\rm s}$, respectively.
In Src.\ B, the longer flare in Obs.\ 4729 is less probable;
however, the error bars of all its inferred properties overlap with the error
bars of the other flares in Src.\ B.
If one removed the Obs.\ 4729
flare, the average flare duration decreases to
$1573^{+1425}_{-1022} \, {\rm s}$.
Although the durations are larger than typical Type-I bursts,
they are consistent within the uncertainties with Type-I superbursts.
Since Type-I bursts are Eddington-limited, observations of distant
galaxies may select for longer bursts, given the small numbers of
photons in each burst.
Thus, we may be observing relatively rare, extreme forms of Type-I
superbursts.

The average luminosities for the flares
in Src.\ A and Src.\ B are
$5.8^{+25.4}_{-\phn2.0}$ and
$2.9^{+26.3}_{-\phn0.5} \times 10^{38} {\rm erg} \, {\rm s}^{-1}$,
respectively, higher than the Eddington luminosity of a $1.4 \, M_\odot$ NS
accreting hydrogen.
If one removed the Obs.\ 4729 flare from Src.\ B, 
the average flare luminosity increases to
$3.8^{+18.4}_{-\phn0.9} \times 10^{38} {\rm erg} \, {\rm s}^{-1}$.
The reported luminosities are for the 0.3--10 keV
band, and the bolometric luminosities would be even higher. Furthermore,
the uncertainties quoted don't include the systematic
uncertainty in the distance to NGC 4697 (although this is smaller
than the statistical uncertainties) or in the spectral model for the
source.
Typically, Galactic Type-I bursts can be modeled as $\sim$2 keV
blackbodies; if this model were used the observed luminosities would increase
by a factor of $\sim 2.7$.

The average fluences of the flares in these two sources are
$4.3^{+3.0}_{-1.6}$ and
$6.0^{+4.0}_{-2.2} \times 10^{41} {\rm erg}$, respectively.
Assuming the flares are powered by fusion of hydrogen
and that the thermonuclear efficiency to the iron-peak elements is
$0.007 c^2$ \citep{LPT1995},
this requires the accretion of $\sim 4 \times 10^{-11} \, M_\odot$.
For accretion of helium and an efficiency of $0.002 c^2$, the mass is
$\sim 1 \times 10^{-10} \, M_\odot$.
The persistent luminosities of the sources are
$\sim 4$ and
$\sim 1 \times 10^{37}  {\rm erg} \, {\rm s}^{-1}$
implying accretion rates of
$\sim 4$ and
$\sim 1 \times 10^{-9} \, M_\odot$ yr$^{-1}$
for an NS.
Thus, it would require days to weeks to build up a layer sufficient to produce
the observed flares as thermonuclear flashes.
As noted above, the recurrence time for the flares is probably
shorter than this, $\Delta T \approx 10 \, {\rm hr}$.
However, if most of the LMXBs in NGC 4697 produce similar flares, then
the only limit is $\Delta T \la 11$ days.
The durations and fluences are similar to superbursts; however,
the intervals between Galactic superbursts are much longer, on order of
a year.

Both sources are close to GCs in NGC 4697; thus, it is unlikely these
sources are background active galactic nuclei (AGNs). Src.\ A is
$0\farcs1$ from an HST-ACS GC with $g-z = 0.89$ and $z=20.91$.  This
source is a highly probable match.  Src.\ B is $0\farcs8$ from an
HST-ACS GC with $g-z = 1.32$ and $z=19.15$.  Within $0\farcs8$ at this
projected position in the galaxy, the probability of a match occurring
at random is 1\%.  Thus, both of these flaring sources are probably
located in GCs.

\subsection{CXOU J124839.0$-$054750 (Src.\ C)}

By itself, the flare in Obs.\ 0784 of Src.\ C is highly
significant. The combination of a similar flare in Obs.\ 4727 makes it
the most probable (99.995\%) flaring source in this Paper.

There is no clear optical counterpart for Src.\ C.
The nearest globular cluster is $1\farcs8$ away. Given the rms of
$0\farcs4$ for X-ray/optical matches, we do not think this is a likely
match.
At this source's radius from the center of
NGC 4697 ($48\arcsec$), one only expects $\sim 0.3$ background
sources at this flux level.
There are 18 other
sources as bright as Src.\ C and interior to it. Therefore,
there is only a 1.6\% probability the source is a background AGN from the
X-ray data alone.
The source is even less likely to be an AGN given the lack of an optical
detection.
We have estimated the 50\% completeness limit near the source to be
24.9 in the $z$-band. Since, the X-ray flux in the
$0.5$--$8.0 \, {\rm keV}$ band is
$4.2\times 10^{-15} \, \rm{erg\,s}^{-1}\,{\rm cm}^{-2}$,
$\log F_{\rm X} / F_{\rm opt}$ is greater than 0.5.
If we relax this to
$\log F_{\rm X} / F_{\rm opt}> 0.25$ ($z > 24.3$), and look at GOODS X-ray
selected sources \citep{TUC+2004}, only 9.6\% of the sources have 
$\log F_{\rm X} / F_{\rm opt} > 0.25$.
Combining the expected X-ray background with the results for obscured AGN in
the GOODS fields, we estimate that there is a 0.15\% probability
Src.\ C is an AGN.

The flare behavior of Src.\ C does not have a clear
analog in our own Galaxy. Its peak luminosity of 
$ 5.5^{+ 11.2}_{- \phn2.3} \times 10^{39} \, \rm{erg\,s}^{-1}$ is
clearly super-Eddington for an NS; the flare is at least 8 times the
Eddington luminosity of a helium burning NS. This shows it is not a
Type-I X-ray burst. Since NGC 4697 is an elliptical galaxy, it is unlikely
that this source is an HMXB like LMC X$-$4 or V4641 Sgr.

The outburst need not exceed the Eddington limit by much if
the compact object is a $10 \, M_{\odot}$ BH. The flare
timescale is similar to rapid transients seen in the BH-XRBs,
GRS1915+105 and V4641 Sgr.
Compared to GRS1915+105, the ratio of flare to persistent
rate for short flares is too high and the recurrence timescale
too long, arguing against accretion behavior like that in GRS1915+105.
The rapid transient in V4641 Sgr, seen at the tail of
a flare event with the
{\it  Rossi X-Ray Timing Explorer}
\citep[RXTE;][]{WK2000},
has the correct duration and peak luminosity
(assuming a distance of $\sim 10 \, {\rm kpc}$, \citealt{OKK+2001});
however, its quiescent luminosity is too low compared to 
Src.\ C by more than a factor of 10.
The activity in V4641 Sgr has
been attributed to super-Eddington accretion onto a black hole
with the formation of an extended envelope \citep{RGC+2002}.
This variability might look like Src.\ C if
it occurred during a long-term luminous stage as has been seen in
GRS1915+105.

Since its behavior is most like LMC X$-$4, one might suggest that they
have a similar phenomenology. That is, the flare might be due to
density inhomogeneities in the accretion columns onto a neutron star
polar caps \citep{MEW2003}. This would require that Src.\ C have a
large magnetic field.
Since most Galactic LMXBs do not appear to
have strong fields, it is thought that magnetic fields in NS binaries
decay with time or as a result of accretion
\citep{BS1995}.
Since a
NS in NGC 4697 is unlikely to be newly formed, a strong magnetic field in a
NS binary could only occur if accretion causes the magnetic field decay
in LMXBs and millisec pulsars, and if
the neutron star in this binary has only recently begun to accrete from
its donor.

One possibility is that Src.\ C (and possibly
Src.\ A \& B as well)
are related to Galactic microquasar sources.
Microquasars are XRBs with accreting BHs which produce
relativistic jets \citep[e.g.,][]{MR1999}.
In most of the known Galactic examples, we are observing the
sources at a large angle from the jet axis 
\citep[see, however,][]{OKK+2001}.
The very high luminosity of Src.\ C might be explained if
we are seeing this source along the jet axis.
In analogy to their AGN counterparts, microquasars observed along the jet
axis are referred to as microblazars \citep{MR1999}.
Blazars are known to undergo relatively short timescale
outbursts; the same phenomena, scaled to microblazars, might
account for the X-ray flares in Src.\ C.

\acknowledgements
We thank Adrienne Juett and Eric Pfahl for very helpful comments.
Support for this work was provided by NASA through HST Award Number
GO-10003.01-A and {\it Chandra} Award Numbers GO4-5093X, AR3-4005X,
GO3-4099X, and AR4-5008X.
G.~R.~S.\ acknowledges the receipt
of an ARCS fellowship and support provided by the F. H. Levinson Fund.

\bibliography{ms}

\end{document}